\documentclass[%
pre,
aps,
a4paper,
english,
showpacs,
reprint,
twocolumn,
superscriptaddress
]{revtex4-2}

\usepackage{epigraph}
\usepackage[T1]{fontenc}
\usepackage[utf8]{inputenc}
\usepackage{amsmath}
\usepackage{amssymb}
\usepackage{graphicx}
\usepackage{refstyle}
\usepackage{skak}
\usepackage{multirow}
\usepackage{epsdice}
\usepackage{soul}
\usepackage[
  citecolor=blue,
  colorlinks,
  linkcolor=blue,
  urlcolor=blue,
]{hyperref}
\usepackage{epstopdf}
\usepackage{bm}
\usepackage[version=4]{mhchem}
\usepackage{tikz}
\usetikzlibrary{arrows,shapes,trees}
\usepackage{booktabs} 
\usepackage{lipsum} 
\usepackage{float}

\usepackage[pagewise]{lineno} 
\begin{document}
	\title{Collaboration drives phase transitions towards cooperation in prisoner's dilemma}
	
	\author{Joy Das Bairagya}
	\email{joydas@iitk.ac.in}
	\affiliation{
		Department of Physics,
		Indian Institute of Technology Kanpur, Uttar Pradesh, PIN: 208016, India
	}
		\author{ Jonathan Newton}
	\email{newton@kier.kyoto-u.ac.jp}
	\affiliation{
		Institute of Economic Research, 
		Kyoto University, Kyoto, Japan
	}
	\author{Sagar Chakraborty}
	\email{sagarc@iitk.ac.in}
	\affiliation{
		Department of Physics,
		Indian Institute of Technology Kanpur, Uttar Pradesh, PIN: 208016, India
	}
\begin{abstract}
{The emergence of cooperation among self-interested individuals is a central puzzle across biology and the social sciences. While evolutionary game theory has identified mechanisms such as spatial structure and group selection to sustain cooperation in the prisoner's dilemma, most models assume individuals revise their strategies independently. In reality, social organisms frequently form temporary coalitions to coordinate their decisions. We introduce a \emph{collaboration ring model} to investigate how such joint strategy updating affects the prisoner's dilemma. Individuals update strategies either alone---choosing a best response---or by forming coalitions where actions change only if every member is at least as well off and at least one is strictly better off. Despite individual incentives favouring defection, we show that cooperation emerges through non-equilibrium phase transitions governed by the propensity to collaborate and the benefit of cooperation. Below a critical threshold, defection dominates; above it, cooperation persists through a continuous or a discontinuous transition. Using pair approximation, we derive phase boundaries that align closely with stochastic simulations. Finally, we demonstrate that these results are robust to variations in neighbourhood and coalition size, suggesting that collaborative decision-making is a general mechanism for sustaining cooperation. Our work has potential connections to both the mutualism phenomenon in non-humans and the shared intentionality phenomenon in humans.}

\end{abstract}

\maketitle
\begin{figure*}
	\centering
	\includegraphics[width=1.0\linewidth]{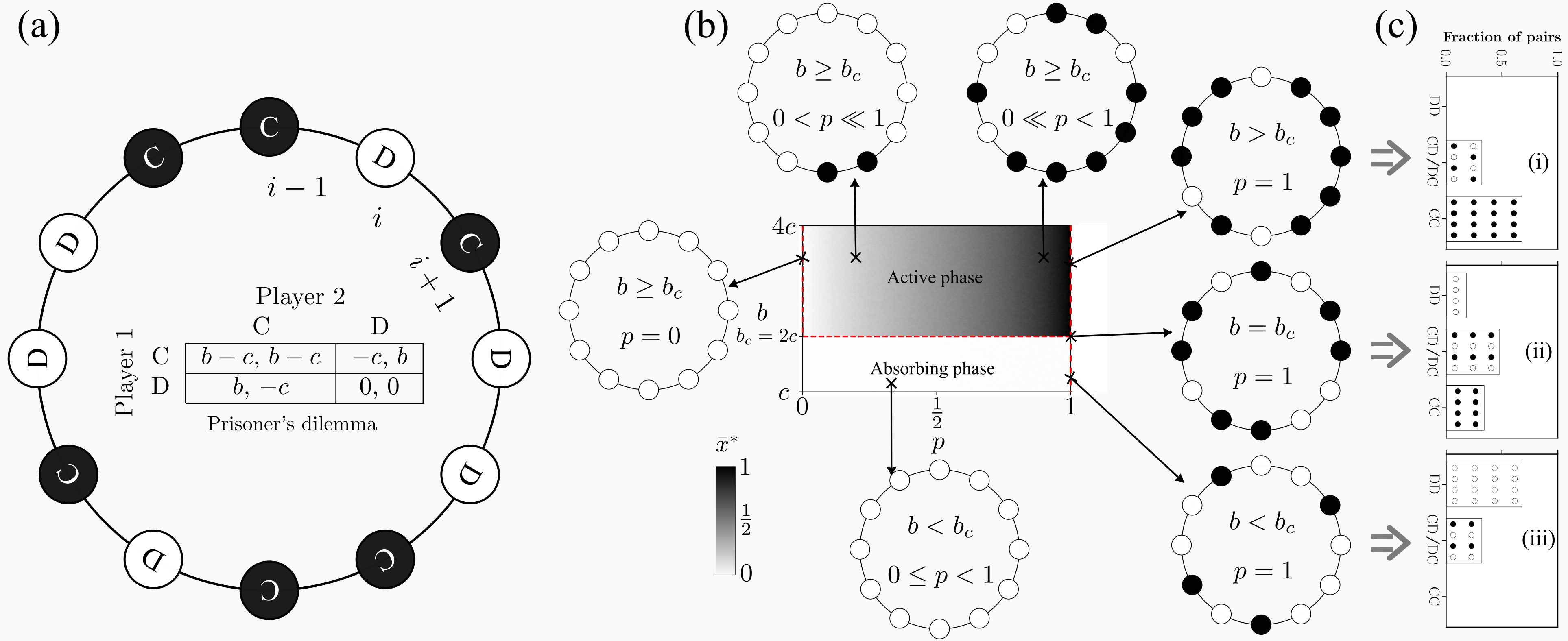}
	\caption{\textbf{Emergence of cooperation, in a PD game, in a ring-structured population through phase transition.} 
	Subfigure (a): Schematic of the ring-structured population: white and black nodes denote defectors and cooperators, respectively, and solid black lines represent nearest neighbour interactions. 
	Subfigure (b): Average order parameter, $\bar{x}^*$, as a function of the control parameters $b$ and $p$, shown using the colour bar. Critical lines where phase transitions occur are indicated by red dashed lines, with illustrative population configuration states shown in different parameter regions. 
	Subfigure (c): Bar plots at $p=1$ display the fractions of nearest-neighbour pairs (DD, CD/DC, CC) for: (i) $b-2c>0$ (only CC and CD/DC present), (ii) $b-2c=0$ (all pair types present), and (iii) $b-2c<0$ (only DD and CD/DC present).}

	\label{fig1}
\end{figure*}
\section{Introduction}
The widespread occurrence of cooperation among non-kin presents a familiar challenge for evolutionary theory. Cooperation among kin is commonly explained through Hamilton's kin selection theory and the associated concept of inclusive fitness~\cite{HAMILTON19641}. By contrast, cooperation among non-kin is often discussed in terms of group or multi-level selection~\cite{MARSHALL2011325,Bourke2011-su,okasha2006evolution,CluttonBrock2009}, as well as a range of individual-level mechanisms such as direct reciprocity, indirect reciprocity, and network reciprocity~\cite{Nowak1992,Nowak1998,Lieberman2005,Ohtsuki2006}. In many such accounts, cooperation is understood as the play of a particular action, often one that is individually costly.

The present paper studies cooperation in the presence of \emph{collaboration}. We retain the standard prisoner’s dilemma distinction between cooperation and defection as primitive actions, but allow neighbouring individuals occasionally to choose their strategies jointly to their mutual benefit. Collaboration is therefore a mutualistic form of strategic choice rather than an altruistic one. The mechanism is immediate and coalitional: it does not rely on repeated interaction or delayed returns, as in direct or indirect reciprocity. We study instead how collaboration can endogenously induce positive local correlation in strategies, and thereby sustain the play of a cooperative action despite unilateral incentives toward defection.

A motivating observation is that examples exist in nature of collaboration by non-kin to generate \emph{immediate} individual benefits. That is, individuals form temporary coalitions to obtain gains that none could secure alone.  Mutualism of this kind is observed across a wide range of biological settings, including bacteria~\cite{HylandKroghsbo2022}, trees~\cite{wohlleben2016hidden}, and animals~\cite{Duguid2020}. A familiar example is avian mobbing: when a predator approaches a nest, nearby birds, often unrelated, may temporarily coordinate in order to harass and drive it away~\cite{Altmann1956,Curio1978,Magrath2014,Pawlak2019}. 

In humans, collaborative behaviour is especially prominent. The literature on shared intentionality studies the ability to form and act on joint intentions, as well as their role in the evolution of human cognition~\cite{Tomasello2010,Tomasello2006,Call2009,vygotsky1980mind,tomasello2014natural,Moll2007}. Our focus is narrower. We study a minimal strategic environment in which collaboration takes the form of temporary coalition formation and joint strategy updating, and ask whether such mutualistic choice can generate cooperation in a prisoner’s dilemma. Our paper therefore contributes to the formal game theoretic study of dynamic processes in the presence of collaboration~\cite{Angus2020,Newton2012,Sawa2014,Newton2015}.

The prisoner’s dilemma is a particularly hostile environment to study whether collaboration can sustain cooperation.  In the game theoretic sense, cooperation is a strictly dominated strategy, so individual best response leads to defection. At the same time, collaboration is constrained by individual rationality: each participant in a coalition must benefit from a joint move. These two features together make the problem non-trivial. A recent numerical study~\cite{Angus2020} showed that, in sparsely connected populations, even small amounts of collaboration can support material and robust levels of cooperation. {The present paper places this insight on a firmer footing by introducing an analytically tractable model. To achieve this, we plan using framework of a ring model that has been shown to be useful for deriving analytically tractable results for the emergence of cooperation~\cite{Ohtsuki2006,Ohtsuki2006_cycle}. Thus, analogously, in this paper, we introduce a ring network and are inspired by an approach used in the literature~\cite{VanBaalen2000,Lieberman2005,Ohtsuki2006rep} to employ an approximation---termed pair approximation---to express the joint probabilities of three and four consecutive ring-sites in terms of nearest-neighbour correlations only, while neglecting all other correlations. This leads us to successfully derive a closed set of solvable deterministic first order ordinary differential equations tracking how collaboration fosters cooperation.} 

Our analysis combines a stochastic local-update process with deterministic approximation techniques. We formulate the underlying dynamics as a non-equilibrium Markov process and study the resulting aggregate behaviour using a pair approximation, related to the Bethe approximation from statistical physics~\cite{1935Bethe,Kikuchi1951,Keeling1999,yedidia2001bethe,Benoit2006}. This allows us to characterize phase transitions  from universal defection to positive cooperation in terms of the propensity to collaborate and the gains generated by cooperation. More broadly, the paper contributes to the literature that uses ideas from statistical physics and dynamical systems to study strategic interaction and collective behaviour~\cite{SZABO20161,szolnoki2004phase,scott2022phase,si2008game,imhof2005long,blume1993statistical,tah2024,Mukhopadhyay2025}.

\section{Collaboration ring model}

Consider a population of $N$ individuals represented as nodes on a ring network~\cite{Knig1990}, for an illustration see Fig.~\ref{fig1}(a). Edges represent interactions between neighbouring individuals. Every individual has two neighbours and plays a prisoner’s dilemma with each of them, using the same strategy in both interactions. The payoff matrix is given in Fig.~\ref{fig1}(a). If a player cooperates, she generates benefit $b$ for her opponent at cost $c$ to herself, where $b>c>0$.

Let $N_C$ denote the number of cooperators and $N_D=N-N_C$ the number of defectors. At each time $\tau\in \mathbb{N}$, an individual $i$ is selected uniformly at random. She then updates her action in one of two ways:
\begin{enumerate}
	\item With probability $p$ (\emph{propensity of collaboration}), she forms a coalition with one of her neighbours, chosen uniformly at random. The coalition members then choose a coalitional better response~\cite{Newton2012,Sawa2014,Newton2015}: they switch jointly to actions that leave each member weakly better off than in the current configuration, with at least one member strictly better off. {In simpler terms, no member should lose any payoff, while at least one member must strictly gain.} If multiple such action profiles exist, one is chosen uniformly at random. If no such action profile exists, both players keep their current actions. \label{better-response}
	
	We emphasize that \emph{collaboration does not necessarily imply cooperation}: a coalition may adopt any joint action profile that satisfies the individual rationality constraints of its members.
	
	\item With probability $1-p$, she updates individually by choosing a best response~\cite{Nash1950} to the fixed actions of her two neighbours. If multiple best responses exist, one is chosen uniformly at random.
\end{enumerate}
{One might alternatively consider stochastic update rules, such as imitation~\cite{SCHLAG1998130}, birth--death~\cite{Moran1958}, or death--birth~\cite{Lieberman2005} processes, as complementary descriptions of the dynamics. However, the primary objective of the present study is to demonstrate that the immediate benefits of temporary coalitions can enhance the probability of cooperation in a population. We therefore deliberately formulate the model without an explicit natural-selection process: everything occurs within a single generation, and each individual can opt for cooperation or defection as they deem appropriate. Furthermore, incorporating imitation would require additional modelling of how individuals of one coalition compare their payoffs with those who belong to a different coalition. If we leave the imitation process aside due to its additional mathematical complexity, constructing a birth--death or death--birth process for the coalitional dynamics is also not straightforward: Since a coalition consists of more than one individual, there is no unique prescription for how the individuals in a coalition should reproduce or whom they should replace, which would introduce unnecessary complexity to the main goal of this paper. Hence, for analytical tractability and to retain the focus on the immediate benefits of temporary collaboration, we restrict our analysis to the best-response and better-response dynamics only.
}

Given this updating rule, our main object of interest is the fraction of cooperators,
\[
x=\frac{N_C}{N}.
\]
Equivalently, $x=\mathrm{P}(\mathrm{C})$, the probability that a randomly chosen node is a cooperator.
 We study how $x$ depends on the benefit of cooperation $b$ and the propensity to collaborate $p$. In the language of statistical physics~\cite{Hauert2005,Szab2007,Szolnoki2011}, $x$ is the order parameter; $b$ and $p$ are control parameters.

\section{Analysis and results}
We analyse the model in two ways. First, we use an agent-based approach to simulate the stochastic process. Second, we derive deterministic dynamics using pair approximation. 
Our aim is to determine whether a positive propensity for collaboration can sustain cooperation in the prisoner’s dilemma, and, if so, how the resulting long-run behaviour depends on the parameters of the model.

\subsection{Agent based dynamics}

We simulate the stochastic Markov process induced by the updating rule described in the previous section. For each time $\tau$, we simulate $R$ independent realizations of the process, all starting with the same cooperator share $x(0)=\frac{1}{2}$. {As $\tau\to\infty$, each realization either becomes trapped in an absorbing state, which we refer to as an \emph{absorbing phase}, or continues to fluctuate according to a limiting distribution, which we refer to as an \emph{active phase}}. Denoting the cooperator fraction in realization $k$ at time $\tau$ by $x_k(\tau)$, we define the ensemble average
\[
\bar{x}(\tau)=\frac{1}{R}\sum_{k=1}^R x_k(\tau).
\]

Fig.~\ref{fig1}(b) illustrates the phase diagram for the long-run ensemble average $\bar{x}^* = \bar{x}(\infty)$ as a function of collaboration propensity $p$ and benefit of cooperation $b$. 
 Along the $x$-axis, we vary $p$ from $0$ to $1$, and along the $y$-axis, we vary $b$ from $c$ to $4c$. For each of $100\times 100$ uniformly spaced points in the $(p,b)$ parameter plane, we simulate $R=500$ independent realizations of length $\tau=10{,}000$, each starting from a random initial configuration with cooperator share $x(0)=\frac{1}{2}$ and population size $N=100$. 
 The value plotted in the figure is then
\begin{equation}
	\bar{x}^* \approx \bar{x}(10{,}000)=\frac{1}{500}\sum_{k=1}^{500}x_k(10{,}000).
\end{equation}

Three classes of long-run behaviour are observed, depending on the value of $p$ and on whether $b$ is greater or less than the critical threshold $b_c\equiv 2c$.
\begin{enumerate}
\item {If $b\leq b_c$ and $p\in[0,1)$, then each realization converges to an absorbing all-defector state, so that $\bar{x}^*=0$. This is also the case when collaboration is absent ($p=0$), even if $b> b_c$.}
\item If $b > b_c$ and $p\in(0,1)$, then the population enters an active phase with $\bar{x}^*\neq 0$. Realizations do not become trapped in absorbing states, and the long-run time average is close to the ensemble average. 
\item {If $p=1$, then the process reaches an absorbing phase, but now there are many possible absorbing states (see Appendix~\ref{absobingstate}).}
\end{enumerate}

These regimes imply three statements about phase transitions. 
First, for fixed $p\in(0,1)$, crossing $b=b_c$ from below we observe a discontinuous transition from all-defection to positive cooperation. This discontinuity is more pronounced at higher values of $p$.
Second, for fixed $b>b_c$, increasing $p$ from $0$ produces a continuous transition away from defection and towards cooperation. 
Finally, at $p=1$ the long-run behaviour changes discontinuously: depending on whether $b<b_c$ or $b>b_c$, the process is absorbed into mixtures of cooperators and defectors with no two adjacent cooperators or no two adjacent defectors, respectively.

\begin{figure}
	\centering
	\includegraphics[width=0.8\linewidth]{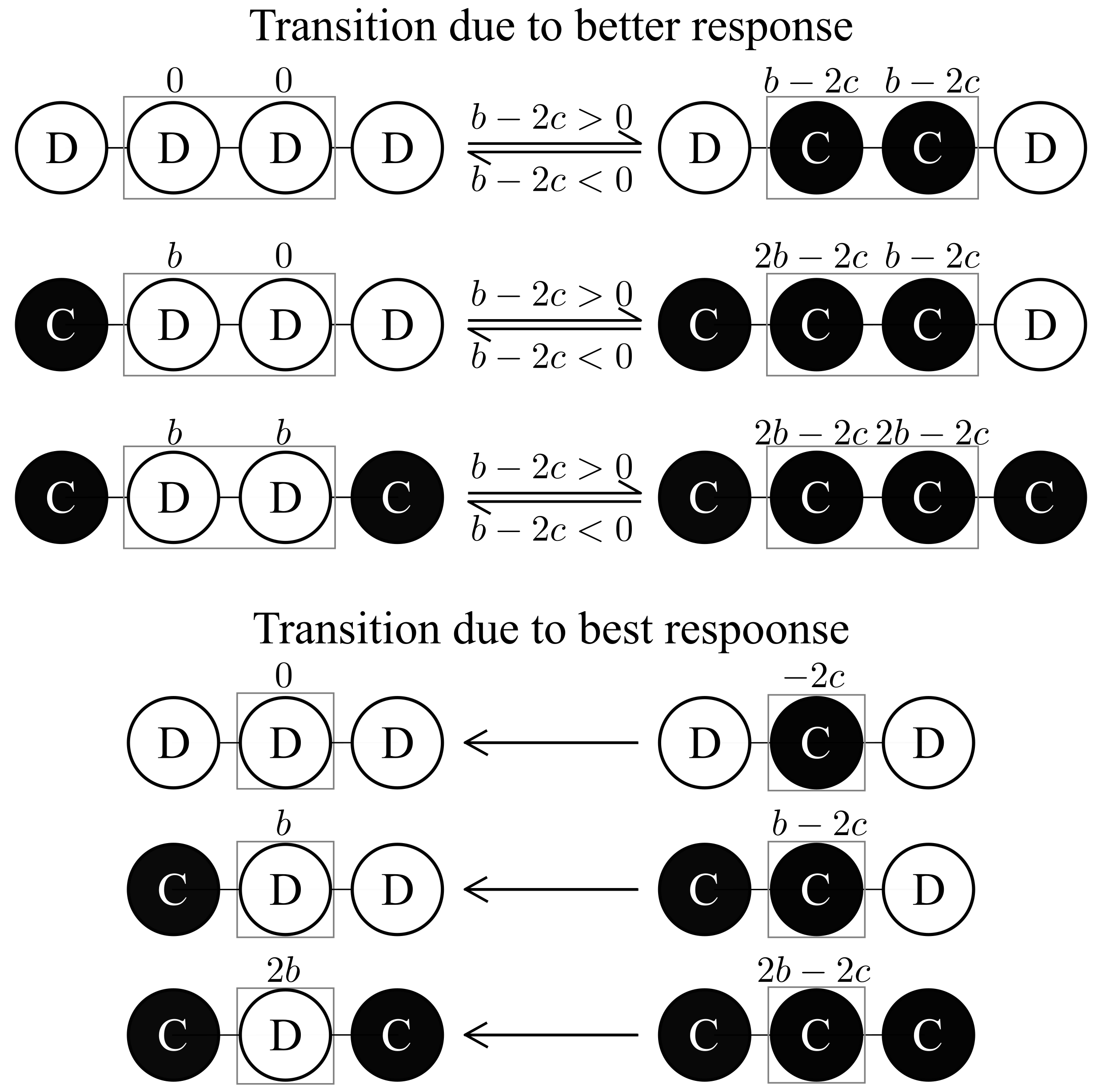}
	\caption{\textbf{All feasible microscopic transitions:} This schematic presents all possible microscopic transitions arising from better-response and best-response updates. In the figure, the arrowheads indicate the direction of change, and above each arrow we specify the parameter range for which the corresponding transition occurs; in case of no specification, the transition is unconditional. { The individual(s), enclosed by a grey rectangular box, is/are chosen to play the game either by forming a coalition or independently. In both the scenarios, we indicate the payoffs they obtain in the given configuration and label their actions inside the circles.} }
	\label{fig2}
\end{figure}

\subsection{Mean-field dynamics with pairwise approximation}
To analyse  the phase transitions identified above, we work from the master equation for the stochastic process~\cite{VANKAMPEN200796}. An exact solution is not available, so we close the resulting hierarchy of equations by approximating the correlations between individuals' actions. The simplest closure is the mean-field approximation~\cite{LANDAU19801}, but this is inadequate in the presence of collaboration, which directly couples the actions of neighbouring individuals (see Sec.~I of the Supplementary Material). We therefore go beyond mean field and adopt a pair approximation, which retains nearest-neighbour correlations while neglecting higher-order ones.

The update rule generates two kinds of transitions: coalitional  and individual. All possible one-step transitions between local configurations are summarized in Fig.~\ref{fig2}. {We emphasize that there is no possibility of a change in the strategies of individuals in a coalition under the better-response scheme when one member of the coalition is a defector and the other is a cooperator. In a PD game, for such a coalition, no other better response exists: in all other altered action configurations, the individual who was defecting would always lose payoff; therefore, there is no better response by definition}.

Since these transitions induce local correlations in actions, we adopt a pair approximation in which correlations are retained only between nearest neighbours. Under this approximation, the state of the population cannot be described solely by the number of cooperators $N_C$ (or defectors {$N_D=N-N_C$)}. One must also specify the number of adjacent pairs of each type. Let $N_{CC}$ and $N_{DD}$ denote the numbers of CC and DD pairs, and let $N_{CD}$ denote the number of mixed pairs, counting both CD and DC. These quantities satisfy
\[
N_{CC}+\frac{N_{CD}}{2}=N_C,\qquad
N_{DD}+\frac{N_{CD}}{2}=N_D.
\]
The first identity follows because every block of $n$ contiguous cooperators contains $n-1$ CC pairs and two mixed boundary pairs. The second identity is similar. A consequence of these identities is that the two variables $N_C$ and $N_{CD}$ suffice {for the {pair-level macroscopic} description of  the population state. We also note that since the range of $N_C$ is $0$ to $N$, the range of $N_{CD}$ is constrained to be $2\min\{N_C,N-N_C\}$ owing to the above two relations among the variables.}

Let ${\rm P}(N_C,N_{CD};\tau)$ be the probability that the population has $N_C$ cooperators and $N_{CD}$ mixed adjacent pairs at time $\tau$. The corresponding master equation is
\begin{widetext}
	\begin{align}
		{\rm P}(N_C, N_{CD}; \tau+1) 
		&= \sum_{\delta_1,\delta_2} 
		{\rm P}\!\left(N_C - \delta_1,\, N_{CD} - \delta_2; \tau\right)\,
		T\!\left(N_C, N_{CD} \mid N_C - \delta_1, N_{CD} - \delta_2\right) \nonumber\\
		&\quad + {\rm P}(N_C, N_{CD}; \tau)
		\left[\,1 - \sum_{\delta_1,\delta_2} 
		T\!\left(N_C + \delta_1, N_{CD} + \delta_2 \mid N_C, N_{CD}\right)\right],
		\label{2dmastereq}
	\end{align}
	where $\delta_1 \in \{-2,-1,+2\}$, $\delta_2 \in \{-2,0,+2\}$, and $T$ denotes the corresponding transition probabilities per unit time.
\end{widetext}

Now, we aim to study the long-term average behaviour of the population. Specifically, we focus on the dynamical evolution of the average fraction of cooperators, $\overline{x}(t)$. From the master equation Eq.~(\ref{2dmastereq}), it follows that the time evolution of $\overline{x}(t)$ cannot be expressed solely in terms of $\overline{x}(t)$, since the transition probabilities given in Table~\ref{tran_c_cd} depend on the strategies of four adjacent individuals. In particular, the dynamics depends on the average fraction of mixed adjacent pairs, ${\overline y} \equiv \overline{\rm P}(C,D) = \frac{\overline{N}_{CD}}{N}$. Therefore, to examine the long-term sustenance of cooperation, one must also consider the dynamical evolution of ${\overline y}(t)$. However, this again requires knowledge of higher-order correlations, namely the average fraction of configurations involving four adjacent individuals. To break this hierarchy, an approximation is necessary.
	
\begin{figure}
	\centering
	\includegraphics[width=1.0\linewidth]{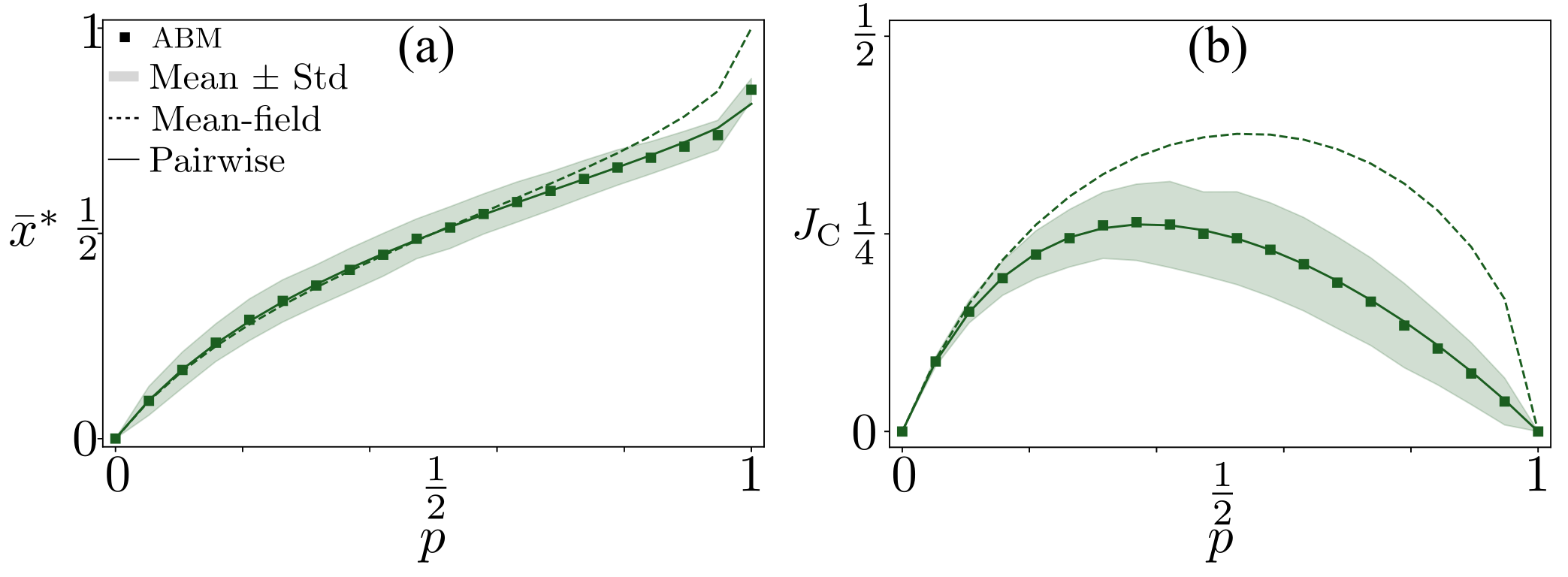}
	\caption{\textbf{Prediction of the average fraction of cooperators and its inflow rate using mean-field and pairwise ansatzes.}
	This figure shows the average fraction of cooperators $\bar{x}^*$ [subfigure (a)], and the inflow rate $J_C$ [subfigure (b)] as functions of the coalition probability $p$ in an active phase with $p<1$, $b = 3$ and $c = 1$, i.e., $b>b_c$, (green colour). Results generated from the numerical simulations are depicted by filled square. All dashed lines denote the mean-field prediction, which deviates significantly from the numerical simulation results. In contrast, the pairwise approximation ansatz, represented by solid line, closely matches the numerical simulation results across the full range of $p$, remaining within one standard deviation of the simulated {ensemble averages.}}
	\label{fig3}
\end{figure}

The simplest approach is the mean-field approximation, which assumes that individuals' strategies are independent. However, this approximation (a detailed analysis has been presented in Section I of Supplementary Material) does not agree with numerical results for higher propensity of collaboration (see Fig.~\ref{fig3}). A more refined approach is the pair approximation ansatz~\cite{Keeling1999,Benoit2006}, which is closely related to the Bethe approximation in statistical physics~\cite{1935Bethe,Kikuchi1951,yedidia2001bethe}.
	
Now, let us introduce the pairwise approximation formally. If we denote the ensemble-averaged probability of observing the local configuration $\alpha_1 \cdots \alpha_k$ on $k$ consecutive nodes by $\overline{{\rm P}}(\alpha_1 \ldots \alpha_k)$, then, under this approximation, one can write:
\begin{subequations}
	\label{eq:pa}
\begin{eqnarray}
\overline{{\rm P}}(\alpha_1\alpha_2\alpha_3\alpha_4)
&\approx&
\overline{{\rm P}}(\alpha_2\alpha_3)\,
\overline{{\rm P}}(\alpha_1\mid\alpha_2)\,
\overline{{\rm P}}(\alpha_4\mid\alpha_3),
\qquad
\label{eq:pa3}\\
\overline{{\rm P}}(\alpha_1\alpha_2\alpha_3)
&\approx&
\overline{{\rm P}}(\alpha_2)\,
\overline{{\rm P}}(\alpha_1\mid\alpha_2)\,
\overline{{\rm P}}(\alpha_3\mid\alpha_2),
\qquad
\label{eq:pa2}\\
\overline{{\rm P}}(\alpha_1\alpha_2)
&\approx&
\overline{{\rm P}}(\alpha_2)\,
\overline{{\rm P}}(\alpha_1\mid\alpha_2),
\qquad
\label{eq:pa1}
\end{eqnarray}
\end{subequations}
where $\alpha_i \in \{\mathrm{C}, \mathrm{D}\}$ for $ i\in\{1,2,3,4\}$. 
Three site probabilities arise from individual updates, whose payoffs depend on a node and its two neighbours, while four site probabilities arise from coalitional updates, whose payoffs depend on an adjacent pair and the neighbours on either side. These are the largest local configurations that enter the one-step transition probabilities, because interactions are nearest neighbour and coalitions have size two.
Using Eq.~(\ref{eq:pa1}) to eliminate the conditional probabilities in Eqs.~(\ref{eq:pa2}) and (\ref{eq:pa3}), the probabilities of three and four site configurations are approximated in terms of one and two site probabilities.

Define the fraction of mixed adjacent pairs by
\[
y\equiv {\rm P}(C,D)=\frac{N_{CD}}{N},
\]
and rescale time by setting $t=\tau/N$. Under the pair approximation and in the limit $N\to\infty$, the master equation reduces to deterministic dynamics of the form
\begin{subequations}
\label{eq:xydt}
\begin{eqnarray}
\frac{d\bar{x}(t)}{dt}&=&f(\bar{x}(t),\bar{y}(t))=J_C-J_D,\\
\frac{d\bar{y}(t)}{dt}&=&g(\bar{x}(t),\bar{y}(t)).
\end{eqnarray}
\end{subequations}
{The explicit forms of $f$ and $g$ have been derived from Eq.~(\ref{2dmastereq}) in Appendix~\ref{pairwise}; see Eqs.~\ref{eq:pair<0}, \ref{eq:pair=0}, and \ref{eq:pair>0}. Here, $J_C = 2\sum_{\delta_2} \overline{T}(N_C+2,N_{CD}+\delta_2 \mid N_C, N_{CD})$, and $J_D = 2\sum_{\delta_2} \overline{T}(N_C-2,N_{CD}+\delta_2 \mid N_C, N_{CD})+\sum_{\delta_2}\overline{T}(N_C-1,N_{CD}+\delta_2 \mid N_C, N_{CD})$, where $\delta_2 \in \{-2,0,+2\}$. Furthermore, $g(\bar{x}(t),\bar{y}(t))=2\sum_{\delta_1} \overline{T}(N_C+\delta_1,N_{CD}+2 \mid N_C, N_{CD})-2\sum_{\delta_1}\overline{T}(N_C+\delta_1,N_{CD}-2 \mid N_C, N_{CD})$, where $\delta_1 \in \{-2,-1,+2\}$.

Here, we want to emphasize that, as $N \to \infty$, the stochastic nature of the system becomes negligible, and the dynamics are governed entirely by the deterministic part. For finite $N$, however, the stochastic dynamics can be described by adding a Gaussian noise term to Eqs.~\ref{eq:xydt}, with zero mean and a standard deviation proportional to $1/\sqrt{N}$. Nevertheless, the deterministic equations derived in the $N \to \infty$ limit can closely capture the time evolution of the mean state of the system, since averaging over the noise term causes its contribution to vanish. In particular, the long-term behaviour of the stochastic system, i.e., its mean steady-state behaviour, can be well approximated by the stable fixed points of these deterministic equations, even though the actual system consists of a finite but large number of individuals. The pair approximation is an independent calculation step only aimed at rendering the deterministic equations analytically solvable by closing the moment hierarchy issue---solving for lower moments, due to the pair approximation, no longer requires the knowledge of the higher moments.}

This deterministic system reproduces the phase structure seen in the simulations. In particular, the steady states of Eq.~(\ref{eq:xydt}) correspond to the phases identified in Fig.~\ref{fig1}(b), and the predicted inflow rate $J_C$ agrees closely with the numerical results. Specifically,

%
\begin{enumerate}
\label{fixed_points}
\item For $b< b_c$ and $p\in[0,1)$, the unique stable fixed point is $(\bar{x}^*,\bar{y}^*)=(0,0)$, corresponding to the absorbing all-defector state. At $p=1$, this fixed point is replaced by a line of fixed points satisfying
\[
\bar{x}^*-\frac{\bar{y}^*}{2}=0.
\]
Each point on this line corresponds to an absorbing state with no CC pairs. Consequently, at $p=1$ the order parameter $\bar{x}^*$ exhibits a discontinuous jump.
	\item {Similarly, for $b=b_c$ and $p\in[0,1)$, the unique stable fixed point is $(\bar{x}^*,\bar{y}^*)=(0,0)$. However for $p =1$, owing to the absence of best and better responses, the dynamics remains {fixed} at the initial condition, any point $(\bar{x}^*,\bar{y}^*)$---where $0\leq\bar{x}^*\leq 1$ and $0\leq \bar{y}^*\leq 2\min\{\bar{x}^*,1-\bar{x}^*\}$---is practically a fixed point.}

\item For $b>b_c$ and $p\in(0,1)$, the unique stable fixed point is
{\footnotesize
\[
\left(
\frac{7p+2-\sqrt{\Delta(p)}}{2(4p+1)},
\quad
\frac{9p^{2}-5p-2+(p+1)\sqrt{\Delta(p)}}{2p(4p+1)}
\right),
\]
}
where $\Delta(p)\equiv -15p^2+12p+4$. In this steady state,
\[
J_C=2p\left(1-\bar{x}^*-\frac{\bar{y}^*}{2}\right).
\]
This fixed point gives the cooperation level in the active phase{, i.e., {there are no absorbing states}, as every state has a finite probability of being revisited. It also matches with the simulated {ensemble averages} closely; see Fig.~\ref{fig3}. At $p=0$, the unique stable fixed point is $(\bar{x}^*,\bar{y}^*)=(0,0)$. At $p=1$, the active phase, and hence the above fixed point, disappears and is replaced by a line of fixed points satisfying $\bar{x}^*+\frac{\bar{y}^*}{2}=1$.
This line corresponds to absorbing states with no DD pairs (for more details one should see the Appendix~\ref{pairwise} in  subsection~\ref{absobingstate}).}
\end{enumerate}
For Fig.~\ref{fig3}, we evaluate 20 uniformly spaced values of $p$ in the interval $[0,1]$. For each value of $p$, we compute the ensemble average $\bar{x}(10{,}000)$ starting from the initial condition $(x_0,y_0)=(0.5,0.5)$, using a population of size $N=100$. For the active phase we set $b=3$ and $c=1$.
{\section{Beyond ring network}
\label{independence}

The idea of possible phase transition due to collaboration that we have presented in this paper through the ring network is rather general. One can argue, as we elaborate in this section, that the phase transition to cooperation state from comparatively more defection-dominated state is possible in arbitrary networks due to better response. 

First we note that whatever be the network structure where PD game is being played between any two nearest neighbours, defection is the unique best response against any action of the opponents. Consequently, no single defector can increase her payoff by unilaterally deviating from defection to cooperation. This further implies that if a coalition (of arbitrary number of players) adopts a response where only a single defector switches her action to cooperation and all other individuals' actions remain unchanged, then the response can not be a better response.

We now examine whether there exists a better-response update that can induce additional cooperators within a coalition that initially consists of non-zero number of defectors. To this end, as illustrated in Fig.~\ref{fig4}, suppose we select a connected set of $m$ nodes (individuals) who form a coalition in a network of players. Note that there may be further edges between the nodes which appear unconnected but, without loss of generality, we are focussing on a focal pair (highlighted in the figure using gray box) and so we are showing only those edges with which they are connected.

Let the degree of the $i$th individual in the coalition be $l_i$ (where $i\in\{1,2,\dots,m\}$) among which $n_i^{C}$ and $n_i^{D}$ are the numbers of cooperators and defectors, respectively. Every two nearest-neighbours are playing the PD game among themselves.

We next consider whether two defectors within the coalition can jointly increase their payoffs by switching to cooperation, while all other individuals' strategies remain unchanged. If these two defectors are not nearest neighbours, then such a simultaneous switching of their strategies to cooperation would not fetch any additional benefit of mutual cooperation to them as they are not playing with each other. Hence, a simultaneous deviation by two non-neighbouring defectors cannot be a better response.

\begin{figure}
	\centering
	\includegraphics[width=1.0\linewidth]{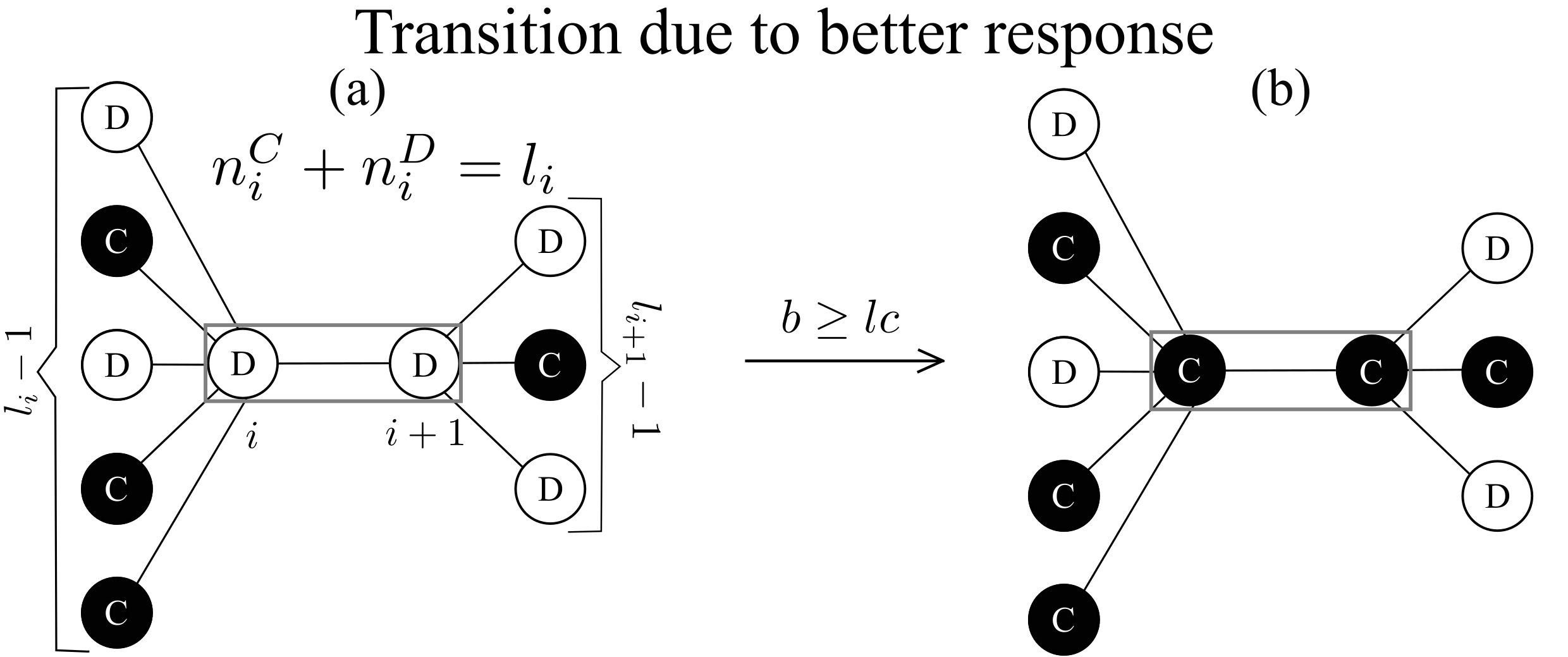}
	\caption{{\textbf{One feasible transition in a network when $b \geq lc$ [see Eq.~(\ref{eq:b_clocal_con})]:} This figure illustrates the transition where two defectors [subfigure (a)] become two cooperators  [subfigure (b)] in a coalition of size $m=10$ in a network driven by better-response dynamics.  Each individual in the coalition is labelled by an index $i \in \{1,\cdots, m\}$. An arbitrary node $j$ has degree $l_j$ and therefore interacts with its $l_j$ nearest neighbours; e.g., in the figure, node $i$ has $l_i=6$ nearest neighbours.}}
		\label{fig4}
	\end{figure}

Next consider the scenario in which two defectors ($i$th and $(i+1)$th members as highlighted by the gray box in Fig.~\ref{fig4}), who are nearest neighbours within the coalition, simultaneously switch to cooperation; all other individuals' strategies are assumed to remain unchanged. We ask whether the action-configuration, in which both switch to cooperation, constitutes a better response, i.e., the payoff of each individual in the new configuration [Fig~\ref{fig4} (b)] must be at least as high as in the original one [Fig~\ref{fig4} (a)], with at least one individual strictly gaining in payoff. Note that for every other member (who has been assumed not to change her action) of the coalition, the simultaneous shift of $i$th and $(i+1)$th individuals from defection to cooperation necessarily results in an increase in her payoff. Specifically, for a defector interacting with either of these individuals, the payoff increases from $0$ to $b$, while for a cooperator the payoff increases from $-c$ to $b-c$. Thus, if the action-switch does not decrease payoff for any of the two focal defectors, the action-configuration in question qualifies as better response strategy. Therefore, mathematically, the conditions for the new configuration [Fig~\ref{fig4} (b)] be a better response are:
\begin{align}
	(n_i^C+1)(b-c) + (n_i^D-1)(-c) & \geq n_i^C b+n_i^D \times 0,\quad \\
	(n_{i+1}^C+1)(b-c) + (n_{i+1}^D-1)(-c) &\geq n_{i+1}^C b+n_{i+1}^D \times 0;\quad
\end{align}
which, respectively, simplify to
\begin{align}
	b -l_ic \geq 0, \\
	b -l_{i+1}c \geq 0,
\end{align}
because $l_i=n_i^C+n_i^D$. These inequalities together imply that a  sufficient condition for such a joint deviation (DD$\to$CC)---along with no change in actions of other collaborators---to be a better response is
\begin{equation}
	b -lc \geq 0,\quad{\rm where},\,\, l \equiv \max\{l_i,\, l_{i+1}\}.\label{eq:b_clocal_con}
\end{equation}
This means there exists a threshold value of $b$, say,
\begin{align}
	b_c=lc,\label{eq:b_clocal}
\end{align}
above which the number of cooperators in the coalition under consideration---and hence in the entire network---increases (by two, to be precise).

The local coalitional structure as exemplified in Fig.~\ref{fig4} can actually be different in different locations of the network. In any locality, thus, one can construct a better response update in a coalition where only two neighbouring defectors change to cooperation simultaneously. The threshold condition, Eq.~(\ref{eq:b_clocal}), is application to any such coalition; only variation comes through the exact value of $l$ that is location dependent.

It follows from Eq.~(\ref{eq:b_clocal}) that a weaker but simpler sufficient condition admitting a better coalitional response can be expressed in terms of the maximum degree of a node in the network, denoted by $l_{\max}$:
\begin{equation}
	b \geq b^*_c \equiv  l_{\max}c,\label{eq:sufficient}
\end{equation} 
where, of course, $b_c\le b^*_c$.
In conclusion, the phase transitions from all defectors to cooperative state are qualitatively a general phenomenon witnessed in all connected networks of players playing PD: one just has to choose sufficient high $b$. Of course, there may be other better response strategies where more than two defectors change to cooperation in the coalition; for more details see Appendix~\ref{app:example}.
}
\section{Discussion}
We have shown that, in a structured population playing the prisoner’s dilemma, collaboration can sustain cooperation once the benefit generated by cooperation exceeds a critical threshold. In this sense, cooperation among non-kin can emerge even in the presence of free riding when coalition formation creates immediate mutualistic gains~\cite{CluttonBrock2009}. Analytically, the paper combines a ring model with a pair approximation that captures the local correlations induced by collaboration, and this approximation reproduces the phase structure and steady-state cooperation levels observed in the simulations.

{The mechanism identified here is not contingent on the ring network structure. In fact, in Sec.~\ref{independence}, we have shown that there exists a critical value of the benefit of cooperation, $b_c$, above which cooperation can be enhanced through better response for any network structure, with an arbitrary number of neighbours ($l_i$) connected to each node and an arbitrary coalition size ($m$). As a consistency check, we observe explicitly that the critical value $b_c$ given in Eq.~(\ref{eq:b_clocal}) recovers the critical point for the ring network derived earlier in this paper: For the collaboration ring model, $l_i=2$ $\forall i$, i.e., $l=2$, leading to $b_c=2c$, which is precisely the critical value at which the all-defector state of the population gives way to a non-zero fraction of cooperators [see Fig.~\ref{fig1} (b)]. } 

In passing, we bring it to the reader's notice that it was reported in the literature~\cite{Angus2020} that collaboration only promotes cooperation in sparser networks (i.e., smaller $l_{\rm max}$)---a conclusion that was based on numerical investigation; no theoretical reason was put forward. Our theoretical analysis can provide {a} reason for this conclusion. But first we must note that the payoff matrix in the above mentioned paper~\cite{Angus2020} is somewhat different; it can be obtained from our payoff matrix in terms of $b'$ and $c'$ using the transformations: $b \to b'$ and $c \to c' - b'$. Observe that the dilemma's condition, $b>c>0$, translates to $c'>b'$ and $c'<2b$. Under this mapping, {Eq.~(\ref{eq:sufficient}) implies that} cooperation emerges only when {$b' \geq b_c'^* \equiv c'\frac{l_{\rm max}}{l_{\rm max} + 1}$}, with the additional constraint that $b'$ is upper bounded by $c'$. Therefore, for a given value of $c'$, the parameter range of $b'$ that supports the active (cooperation) phase is {$c' - b_c'^* = \frac{c'}{l_{\rm max} + 1}$}. This implies that in sparser networks (smaller $l_{\rm max}$), the parameter range supporting cooperation is larger, and hence cooperation is more readily observed in PD games with payoff matrix in terms of $b'$ and $c'$~\cite{Angus2020}. In our case, however, cooperation always emerges in the presence of collaboration, even for larger values of $l_{\rm max}$; although the critical threshold $b_c$ above which cooperation emerges increases with $l_{\rm max}$.

Apparently, our threshold condition is reminiscent of results from network reciprocity, in that the critical benefit-to-cost ratio depends on local degree~\cite{Ohtsuki2006,Nowak2006}. However, one should recall that the promotion of cooperation via network reciprocity depends on the specific microscopic updating rule; in particular, under death--birth updating, the emergence of cooperation is more likely than under neutral drift. In contrast, collaboration-driven cooperation is fostered via better response dynamics. Moreover, the condition for the emergence of cooperation among non-kin through collaboration is mathematically similar to the condition for cooperation among kin, i.e., Hamilton's kin selection rule~\cite{HAMILTON19641}: $\frac{b}{c} > \frac{1}{r}$, where $r$ is the relatedness. Therefore, collaboration-induced cooperation can also be interpreted as an emergent analogue of kin selection---in the former, network structure that harbours better response dynamics replaces genetic relatedness as the key determinant of cooperative behaviour.

Our model combines ideas from non-cooperative game theory and cooperative game theory through dynamics at both the individual and coalitional levels. Importantly, collaboration should not be confused with group selection: players do not optimize group-level outcomes, and collaboration is not altruistic. Rather, individuals switch jointly only when doing so improves each coalition member’s payoff.
As in network reciprocity or group selection, cooperation ultimately gains from positive local assortment: cooperative strategies perform well in sufficiently concentrated clusters despite losses at their boundaries. What is distinctive here is that collaboration intentionally generates the relevant local assortment.

{A key limitation of the model is its assumption that collaboration is costless. Introducing a cost---borne either by all coalition members or solely by the initiator---would likely shorten the cooperative phase and could eliminate it once the benefits of coordination no longer outweigh the cost. The location of the phase transition would then depend on the cost, the temptation parameter, and coalition frequency. Communication introduces a related challenge. When it is costless but unverifiable, it may generate cheap talk~\cite{Crawford1982,WARNERYD1993532}, deception, or the exploitation of information. When communication is costly, by contrast, it may function as a form of signalling, provided that the cost is difficult for deceptive individuals to imitate. Future work could therefore examine whether honest collaboration can be stabilized through costly signalling, thereby relaxing the current assumption that coalition members commit to their jointly chosen strategy.}

Several other natural extensions follow from this work. There exists a literature on collaboration in coordination problems and convention dynamics~\cite{Newton2015,newton2019watercooler,newton2020agency,newton2021heterogeneity}. A natural next step would be to bring the analytical tools used here, in particular deterministic approximation, to those settings. Furthermore, it is documented that the tendency to live in groups or to adopt a solitary lifestyle has a genetic basis~\cite{Wang2013,Muers2013,Robinson2005}. Indeed, there is a growing literature on the evolution of collaboration itself~\cite{AngusNewton2015PLOS,newton2017shared,Rusch2019,newton2025maxmin}. Here too, it would be valuable to understand whether methods of the kind used in the present paper can yield tractable aggregate dynamics and sharper comparative statics. Finally, the framework motivates behavioural experiments on both human and non-human animals to examine how collaboration influences strategic outcomes, and to test whether our theoretical predictions qualitatively capture behavioural patterns.

\section*{Data availability statement}
All codes used to generate the plots in this paper are available at Zenodo:~\href{https://doi.org/10.5281/zenodo.22913017}{https://doi.org/10.5281/zenodo.22913017}. \\
\acknowledgements
J.D.B. gratefully acknowledges IIT Kanpur (India) for the financial support through Fellowship for Academic and Research Excellence (FARE). Khushi is thanked for validating mathematical calculations done in this paper. This research was supported in part by the International Centre for Theoretical Sciences (ICTS) for the program Decisions, Games, and Evolution (code: ICTS/DGE2025/03). 
\appendix
\section{Thermodynamic limit with pairwise approximation}
\label{pairwise}
From Eq.~(2) of the main text, one can write the following equations 
\begin{widetext}
	\begin{align}
		\overline{\Delta N_C(\tau)}&=\overline{N_C}(\tau+1)-\overline{N_C}(\tau)={\sum_{N_{CD}=0}^N\sum_{N_C=0}^N N_C\left[{\rm P}(N_C,N_{CD};\tau+1)-{\rm P}(N_C,N_{CD};\tau)\right]}.\\
		\overline{\Delta N_{CD}(\tau)}&=\overline{N_{CD}}(\tau+1)-\overline{N_{CD}}(\tau)={\sum_{N_{CD}=0}^N \sum_{N_C=0}^NN_{CD}\left[{\rm P}(N_C,N_{CD};\tau+1)-{\rm P}(N_C,N_{CD};\tau)\right]}.\label{eq:nbar_pair}
	\end{align}
	From the above equations {(for the intermediate steps, one should refer to Eq.~3--9 in the Supplementary Material)}, it is {straightforward} to write the changes in $N_C$ and $N_{CD}$ as follows:
	\begin{align}
		\overline{\Delta N_C(\tau)}&=\sum_{\delta_1,\delta_2} \delta_1 {{\overline T}}\!\left(N_C + \delta_1, N_{CD} + \delta_2 \mid N_C, N_{CD}\right)\\
		\overline{\Delta N_{CD}(\tau)}&=\sum_{\delta_1,\delta_2} \delta_2 {{\overline T}}\!\left(N_C + \delta_1, N_{CD} + \delta_2 \mid N_C, N_{CD}\right),\label{eq:pair_change}
	\end{align}
\end{widetext}
where $\delta_1 \in \{-2,-1, +2\}$, $\delta_2 \in \{-2, 0, +2\}$, and $T$ (given in Table~\ref{tran_c_cd}) denotes the respective transition probabilities per unit time.

Therefore, to write the differential equations for $\bar{x}(\tau)$ and $\bar{y}(\tau)$, at the first step, we have to rewrite the microscopic transition rules which account for the macroscopic changes, i.e., changes in both $N_{C}$ and $N_{CD}$, in the population. All relevant transition probabilities are given in Table~\ref{tran_c_cd}.
\begin{table*}[htbp]
	\centering
	\scriptsize
	\renewcommand{\arraystretch}{1.35}
	\begin{nolinenumbers}
	\resizebox{0.8\textwidth}{!}{%
		\begin{tabular}{|p{3cm}|p{2.5cm}|p{2.5cm}|p{2.5cm}|}
			\hline
			\textbf{Transition} 
			& $\mathbf{b - 2c < 0}$ 
			& $\mathbf{b - 2c = 0}$ 
			& $\mathbf{b - 2c > 0}$ 
			\\ \hline \hline
			
			$T(N_C+2,N_{CD}+2)$
			& $0$
			& {$0$}
			& $p\,\mathrm{P}({}{\rm DDDD})$
			\\ \hline
			
			$T(N_C-2,N_{CD}+2)$
			& $p\,\mathrm{P}({}{\rm CCCC})$
			& {$0$}
			& $0$
			\\ \hline
			
			$T(N_C+2,N_{CD})$
			& $0$
			& {$0$}
			& $p\,\mathrm{P}({}{\rm DDDC})+ p\,\mathrm{P}({}{\rm CDDD})$
			\\ \hline
			
			$T(N_C-2,N_{CD})$
			& $p\,\mathrm{P}({}{\rm CCCD}) + p\,\mathrm{P}({}{\rm DCCC})$
			& {$0$}
			& $0$
			\\ \hline
			
			$T(N_C+2,N_{CD}-2)$
			& $0$
			& {$0$}
			& $p\,\mathrm{P}({}{\rm CDDC})$
			\\ \hline
			
			$T(N_C-2,N_{CD}-2)$
			& $p\,\mathrm{P}({}{\rm DCCD})$
			& {$0$}
			& $0$
			\\ \hline
			
			$T(N_C-1,N_{CD}+2)$
			& $(1-p)\mathrm{P}({}{\rm CCC})$
			& $(1-p)\mathrm{P}({}{\rm CCC})$
			& $(1-p)\mathrm{P}({}{\rm CCC})$
			\\ \hline
			
			$T(N_C-1,N_{CD})$
			& $(1-p)\mathrm{P}({}{\rm CCD})  +  \,\,\,\,\,\,(1-p)\mathrm{P}({}{\rm DCC})$
			& $(1-p)\mathrm{P}({}{\rm CCD})  +  \,\,\,\,\,\, (1-p)\mathrm{P}({}{\rm DCC})$
			& $(1-p)\mathrm{P}({}{\rm CCD})  +  \,\,\,\,\,\, (1-p)\mathrm{P}({}{\rm DCC})$
			\\ \hline
			
			$T(N_C-1,N_{CD}-2)$
			& $(1-p)\mathrm{P}({}{\rm DCD})$
			& $(1-p)\mathrm{P}({}{\rm DCD})$
			& $(1-p)\mathrm{P}({}{\rm DCD})$
			\\ \hline
		\end{tabular}}%
		\end{nolinenumbers}
	\caption{{Transition probabilities per unit time: The table summarizes all possible transition probabilities occurring in a single update step, $\tau \to \tau+1$, from the state $(N_C,N_{CD})$. For brevity, we use the shorter notation for $T(N_C+\delta_1,N_{CD}+\delta_2\mid N_C,N_{CD})$ as $T(N_C+\delta_1,N_{CD}+\delta_2)$ in the table.} The expression $\mathrm{P}\!\left({}{\alpha_1, \alpha_2, \ldots, \alpha_{\kappa}}\right)$ denotes the joint probability of the action sequence $\alpha_1, \alpha_2, \ldots, \alpha_{\kappa}$, where each consecutive pair $\alpha_i$ and $\alpha_{i+1}$ is connected, and the terminal actions $\alpha_1$ and $\alpha_{\kappa}$ remain fixed. Each $\alpha_i$ belongs to $\{\mathrm{C}, \mathrm{D}\}$, and the sequence length $\kappa$ is either $3$ or $4$.	
	}
	\label{tran_c_cd}
\end{table*}

Making use of Table.~(\ref{tran_c_cd}) and pair approximation [main text Eq.~4] in Eq.~(\ref{eq:nbar_pair}); subsequently, rescaling time by defining \(t=\tau/N\), one finds in the limit of large \(N\) (so that \(1/N \to 0\)), Eq.~(\ref{eq:nbar_pair}) becomes for $b - 2c < 0$, $b - 2c = 0$ and $b - 2c > 0$, respectively,
\begin{enumerate}
	
	\item {For $b - 2c < 0$:}
	\begin{subequations}\label{eq:pair<0}
		\begin{align}
			\frac{d\,\bar{x}(t)}{dt} 
			&= -2p\left(\bar{x}(t) - \frac{\bar{y}(t)}{2}\right) - (1-p)\bar{x}(t), \\
			\frac{d\,\bar{y}(t)}{dt}
			&\approx 
			2\left(\frac{\bar{x}(t) - \bar{y}(t)}{\bar{x}(t)}\right)
			\left[\bar{x}(t) - p\frac{\bar{y}(t)}{2}\right].
		\end{align}
	\end{subequations}
	
	\item {For $b - 2c = 0$:}
	{\begin{subequations}\label{eq:pair=0}
		\begin{align}
			&\frac{d\,\bar{x}(t)}{dt}= 0 - (1-p)\bar{x}(t), \\
			&\frac{d\,\bar{y}(t)}{dt}\approx 2(1-p)\left[\bar{x}(t)-\bar{y}(t)\right].
		\end{align}
	\end{subequations}}
	
	\item {For $b - 2c > 0$:}
	\begin{subequations}\label{eq:pair>0}
		\begin{align}
			\frac{d\,\bar{x}(t)}{dt}
			&= 2p\left(1 - \bar{x}(t) - \frac{\bar{y}(t)}{2}\right)
			- (1-p)\bar{x}(t), \\
			\frac{d\,\bar{y}(t)}{dt}
			\approx&
			2p\left[
			\frac{
				[1 - \bar{x}(t) - \bar{y}(t))\,
				[1 - \bar{x}(t) - \frac{\bar{y}(t)}{2})
			}{
				1 - \bar{x}(t)
			}
			\right]
			\nonumber \\
			&\quad +
			2(1-p)\big(\bar{x}(t) - \bar{y}(t)\big).
		\end{align}
	\end{subequations}
\end{enumerate}
{All the steps are exactly analogous to those in the mean-field approximation, which are provided in Eqs.~(3)--(13) of the Supplementary Material.}
\subsection{Many absorbing states at $p=1$}
\label{absobingstate}
{At $p = 1$, the dynamics is governed solely by better-response updates. In this regime, for $b - 2c < 0$, all CC pairs are converted into DD pairs (see Fig.~\ref{fig2}), and the system evolves towards absorbing states characterized by the absence of CC pairs, i.e., $N_{CC} = 0$. Conversely, for $b - 2c > 0$, all DD pairs are converted into CC pairs (see Fig.~\ref{fig2}), and the system reaches absorbing states in which no DD pairs remain, i.e., $N_{DD} = 0$. Below we have shown that the mean-field dynamical predictions in these regimes are in full agreement with the expected stable fixed points.}

All the stable fixed points of the set of differential equations, for three parameter ranges of $b$, are given in the main text. Here, we want to discuss the two sets of non-isolated stable fixed points that arise at $p=1$ when $b \neq b_c$. 
\begin{enumerate}
	\item {For $b - 2c < 0$ and $p=1$:}   
	In this parameter range, the stable set of non-isolated fixed points fall on a line whose equation is given {below}:
	\[
	\bar{x} - \frac{\bar{y}}{2} = 0.
	\]
	This line corresponds to an absorbing state of an infinite population and represents the absence of CC pair. Now let us determine for a finite population, how many microstates exist such that $N_{CC}=0$ and distinguishable by their macroscopic description, i.e., $(N_{DD}, N_{CD})$. 
	For finite $N$, the exact number of distinct absorbing states at $p=1$ can be computed by counting the integer pairs $(N_{DD}, N_{CD})$ satisfying the constraint
	\[
	N_{DD} + N_{CD} = N.
	\]
	Although this yields $N+1$ possible pairs, $N_{CD}$ must be even, reducing the number of valid absorbing states to  
	$\frac{N+1}{2}$ $\text{if $N+1$ is even}$ and $\frac{N+2}{2}$ $\text{if $N+1$ is odd}$.
	
	\item {For $b - 2c > 0$ and $p=1$:}  
	Similar to the above discussed  case, for this parameter range, the stable  set of non-isolated fixed points are
	$
	\bar{x}^* + \frac{\bar{y}^*}{2} = 1
	$. This line corresponds to the absorbing state in which no DD pair exists.
	
	Similar to the previous case, here also let us determine the exact number of distinct absorbing states at $p=1$ for finite $N$. The number of distinct absorbing states can be computed by counting the integer pairs $(N_{CC}, N_{CD})$ satisfying the constraint
	\[
	N_{CC} + N_{CD} = N.
	\]
	Although this yields $N+1$ possible pairs, $N_{CD}$ must be even, reducing the number of valid absorbing states to  
	$\frac{N+1}{2}$ $\text{if $N+1$ is even}$ and $\frac{N+2}{2}$ $\text{if $N+1$ is odd}$. 

\end{enumerate}

{\section{Other better responses}
\label{app:example}
We ask: Are there coalitional better responses where more than two defectors in the coalition change their actions simultaneously to cooperation? The answer is in affirmative as can be easily seen below. 

Suppose there is a set $\mathcal{D}$ of more than two, say {$\eta^D (<m)$}, connected defectors so that $i$th defector in $\mathcal{D}$ has $\eta^D_i\in\{1,2,\cdots,\eta^D-1\}$ nearest neighbours in $\mathcal{D}$. The player (whether a cooperator or a defector) who is not in $\mathcal{D}$ (but is in the coalition) will be better off when all the defectors in $\mathcal{D}$ turn to cooperators. Therefore, along the line of arguments already given for the case $\eta^D=2$, switching of action to cooperation by all the defectors in $\mathcal{D}$ (along with the unchanged action of the players not in $\mathcal{D}$) would constitute a better response if
\begin{align}
		&(n_i^C+\eta^D_i)(b-c) + (n_i^D-\eta^D_i)(-c) \geq n_i^C b+n_i^D \times 0\qquad\\
		\implies& l_i(-c) + \eta^D_ib \geq 0,
\end{align}
$\forall i\in\mathcal{D}$. As before, this means there exists a threshold value of $b$, say,
\begin{align}
	b_{\rm th}=c\,\frac{\max_{i\in\mathcal{D}} l_i}{\min_{i\in\mathcal{D}} \eta^D_i},\label{eq:b_clocal1}
\end{align}
above which cooperators increase in the network due to the better response.

Since many such sets like $\mathcal{D}$ can exist, the sufficient value of $b$, i.e., the threshold $b_{\rm th}^{\max}$ beyond which cooperation can be fostered by better response using arbitrary $\mathcal{D}$ can be expressed as
\begin{equation}
	b_{\rm th}^{\max}=	c\,\max_{\mathcal{D}}
	\left(\frac{\max_{i\in\mathcal{D}} l_i}{\min_{i\in\mathcal{D}} \eta^D_i}\right).
\end{equation}
}

\bibliography{Bairagya_col}

\end{document}